\documentclass[12pt]{article}
\usepackage{graphicx}

\newcommand\pubnumber{SLAC--PUB--14481}
\newcommand\pubdate{June, 2011}

\def\SLAC{SLAC, 
    Stanford University, Menlo Park, CA 94025 USA}
\def\doeack{\footnote{Work supported by the US Department of Energy,
                     contract DE--AC02--76SF00515.}}

\def\Title#1{\begin{center} {\Large #1 } \end{center}}
\def\Author#1{\begin{center}{ \sc #1} \end{center}}
\def\Address#1{\begin{center}{ \it #1} \end{center}}

\def\submit#1{\begin{center}Submitted to {\sl #1} \end{center}}
\newcommand\pubblock{\rightline{\begin{tabular}{l} \pubnumber\\
         \pubdate \end{tabular}}}
\newenvironment{Abstract}{\begin{quotation} \begin{center}
                       ABSTRACT
     \end{center}\bigskip  }{\end{quotation}}

\def\submit#1{\begin{center}Submitted to {\sl #1} \end{center}}
\def\Acknowledgements{\bigskip  \bigskip \begin{center} \begin{large}
             \bf ACKNOWLEDGEMENTS \end{large}\end{center}}

\def\beq{\begin{equation}}
\def\eeq#1{\label{#1}\end{equation}}
\def\eeqn{\end{equation}}

\newenvironment{Eqnarray}%
   {\arraycolsep 0.14em\begin{eqnarray}}{\end{eqnarray}}
\def\beqa{\begin{Eqnarray}}
\def\eeqa#1{\label{#1}\end{Eqnarray}}
\def\eeqan{\end{Eqnarray}}
\def\CR{\nonumber \\ }

\def\leqn#1{(\ref{#1})}

\let\bar=\overbar

\def\bra#1{\left\langle{ #1} \right|}
\def\ket#1{\left| {#1} \right\rangle}

\def\lsim{\mathrel{\raise.3ex\hbox{$<$\kern-.75em\lower1ex\hbox{$\sim$}}}}
\def\gsim{\mathrel{\raise.3ex\hbox{$>$\kern-.75em\lower1ex\hbox{$\sim$}}}}

\def\M{{\cal M}}

\def\tr{{\mbox{\rm tr}}}
\def\half{\frac{1}{2}}
\def\thalf{\frac{3}{2}}

\def\del{\partial}
\def\Dslash{\not{\hbox{\kern-4pt $D$}}}
\def\dslash{\not{\hbox{\kern-2pt $\del$}}}

\def\CM{{\mbox{\scriptsize CM}}}

\def\msb{{\bar{\scriptsize M \kern -1pt S}}}

\def\drb{{\bar{\scriptsize D \kern -1pt R}}}

\def\eps{\epsilon}

\def\spa#1#2{\langle #1 #2 \rangle}
\def\spb#1#2{[ #1 #2 ]}
\def\apb#1{\langle #1  ]}
\def\bpa#1{[ #1  \rangle}

\makeatletter
\def\section{\@startsection{section}{0}{\z@}{5.5ex plus .5ex minus
 1.5ex}{2.3ex plus .2ex}{\large\bf}}
\def\subsection{\@startsection{subsection}{1}{\z@}{3.5ex plus .5ex minus
 1.5ex}{1.3ex plus .2ex}{\normalsize\bf}}
\def\subsubsection{\@startsection{subsubsection}{2}{\z@}{-3.5ex plus
-1ex minus  -.2ex}{2.3ex plus .2ex}{\normalsize\sl}}

\renewcommand{\@makecaption}[2]{%
   \vskip 10pt
   \setbox\@tempboxa\hbox{\small #1: #2}
   \ifdim \wd\@tempboxa >\hsize     
       \small #1: #2\par          
     \else                        
       \hbox to\hsize{\hfil\box\@tempboxa\hfil}
   \fi}

 \def\citenum#1{{\def\@cite##1##2{##1}\cite{#1}}}
 
\newcount\@tempcntc
\def\@citex[#1]#2{\if@filesw\immediate\write\@auxout{\string\citation{#2}}\fi
  \@tempcnta\z@\@tempcntb\m@ne\def\@citea{}\@cite{\@for\@citeb:=#2\do
    {\@ifundefined
       {b@\@citeb}{\@citeo\@tempcntb\m@ne\@citea\def\@citea{,}{\bf ?}\@warning
       {Citation `\@citeb' on page \thepage \space undefined}}%
    {\setbox\z@\hbox{\global\@tempcntc0\csname b@\@citeb\endcsname\relax}%
     \ifnum\@tempcntc=\z@ \@citeo\@tempcntb\m@ne
       \@citea\def\@citea{,}\hbox{\csname b@\@citeb\endcsname}%
     \else
      \advance\@tempcntb\@ne
      \ifnum\@tempcntb=\@tempcntc
      \else\advance\@tempcntb\m@ne\@citeo
      \@tempcnta\@tempcntc\@tempcntb\@tempcntc\fi\fi}}\@citeo}{#1}}
\def\@citeo{\ifnum\@tempcnta>\@tempcntb\else\@citea\def\@citea{,}%
  \ifnum\@tempcnta=\@tempcntb\the\@tempcnta\else
  {\advance\@tempcnta\@ne\ifnum\@tempcnta=\@tempcntb \else\def\@citea{--}\fi
    \advance\@tempcnta\m@ne\the\@tempcnta\@citea\the\@tempcntb}\fi\fi}
\makeatother

\def\afl{{a^\flat}}
\def\ash{{a^\sharp}}
\def\bfl{{b^\flat}}
\def\bsh{{b^\sharp}}
\def\cfl{{c^\flat}}
\def\csh{{c^\sharp}}
\def\Afl{{A^\flat}}

\def\Bfl{{B^\flat}}

\begin{document}
\begin{titlepage}
\pubblock

\vfill
\Title{Antenna Splitting Functions for Massive Particles}
\vfill
\Author{Andrew J. Larkoski and Michael E. Peskin\doeack}
\Address{\SLAC}
\vfill
\begin{Abstract}
An antenna shower is a parton shower in which the basic move is a 
color-coherent $2\to 3$ parton splitting process.  In this paper, we 
give compact forms for the spin-dependent antenna splitting functions
involving massive partons of spin 0 and spin $\half$.  
\end{Abstract}
\vfill
\submit{Physical Review {\bf D}}
\vfill
\end{titlepage}
\def\thefootnote{\fnsymbol{footnote}}
\setcounter{footnote}{1}
\tableofcontents
\newpage
\setcounter{page}{1}

\section{Introduction}

The modeling of physics at high energy colliders relies heavily on 
our understanding of QCD.   Quarks and gluons -- collectively, partons -- 
 that are produced
in high energy reactions are observed as jets of hadrons.  The structure
of each jet is determined by the pattern of 
radiation of additional partons from the original one produced in the central 
hard scattering reaction. For this reason, much attention has been given the 
past few years to the development of methods for creating parton showers,
systems of partons created with the distributions predicted by QCD.

The traditional approach to the generation of parton showers is based on 
splitting off partons through a $1\to 2$ branching process.  This 
philosophy is incorporated in the widely used event generator programs
PYTHIA~\cite{PYTHIA} and HERWIG~\cite{HERWIG}.  The construct of building
a shower from $1\to 2$
branching, often called  a `dipole shower', omits an important aspect of the
physics.  The longitudinal momentum distribution in the $1\to 2$ splitting
is given by the Altarelli-Parisi splitting functions~\cite{AP}. 
 In QCD, partons are emitted coherently from the two legs of a 
color dipole.  The emission amplitude is then enhanced inside the dipole and,
more importantly, cancels outside the dipole.  In the 1980's, 
Marchesini and Webber argued that this effect could be incorporated into 
dipole showers by imposing angular ordering of emissions~\cite{MW}.
Thus, HERWIG is built around an angular-ordered parton shower, and PYTHIA,
though it uses a different ordering scheme to choose its branchings, 
vetos emissions that are out of angular ordering.

Alternatively, one might build up a parton shower directly from the color
dipoles, using the $2\to 3$ process of emission of a parton by a dipole as
the basic branching process. This construct is called 
an `antenna shower'. The scheme was realized in the 
program ARIADNE, by Andersson, Gustafson, L\"onnblad, and 
Pettersson~\cite{ARIADNE} and, more recently, by the program VINCIA,
by Giele, Kosower, and Skands~\cite{VINCIA}.   The approach is of 
interest both in creating new parton shower codes for the purpose of
matrix element-parton shower matching and because of its promise to yield
 a more accurate treatment of color dynamics in parton showers.   

Recently,
there has been much interest in the tagging of boosted heavy particles
such as the top and Higgs observed as exotic jets~\cite{tagging}.  Since 
tagging methods rely heavily on color flow, it is interesting to have 
a variety of approaches to the simulation of color flow in parton 
showers in order to test the robustness of these algorithms.

We have been engaged in providing a well-defined foundation for 
antenna showers, giving explicit calculations of the splitting 
functions that generate these showers and generalizing previous work
to spin-dependent formulae.  In a previous paper, we presented the 
complete set of spin-dependent antenna splitting functions needed to 
describe quark and gluon parton showers~\cite{LP}.  In this paper, we 
continue our study of this approach by presenting the spin-dependent antenna
 splitting functions 
for showers with massive particles.   In constructing a shower for 
massless particles, spin-dependence is a convenience, especially for 
matching with full QCD amplitudes.  For massive particles, it is more
important to preserve spin information, because the
decays of heavy particles such as the top quark are spin-dependent and so
the experimental acceptance for the heavy particles
varies significantly with their longitudinal polarization.

The formalism presented here has the same strengths and weaknesses as our 
previous work.  We will calculate in the kinematics of final-state 
showers, using effective operators of definite spin to represent the 
2-particle color dipole state before the splitting.  We will work 
in the limit of a large number of colors in QCD for which the 
concept of a color dipole is strictly defined.  Within this approximation,
 we will derive formulae for splitting functions with any
ratio $m/Q$ between the mass of the particle and the mass of the two-particle
system.
These formulae will necessarily be less simple than those found in~\cite{LP}
for the massless case.  We will see, though, that we can make use of 
spinor product formalism~\cite{spinorintro}
 to write these splitting functions relatively
compactly.  The simplicity of these expressions is connected to their
relation to the Maximally Helicity Violating
 amplitudes of QCD.  This point was originally made for the massless
case in~\cite{Maltoni} and is discussed in some detail in~\cite{LP}. 

The formalism of QCD antennae was originally developed as a tool for 
the subtraction of infrared divergences in higher-order QCD 
calculations.  This approach to QCD calculation was pioneered by 
by Kosower~\cite{Davidone,Davidtwo}.  Gehrmann-De Ridder,
Gehrmann, Glover, and their students have developed this approach
into a sophisticated method applicable to NLO and even NNLO 
computations~\cite{Gehrmann,Daleo}.   Using this formalism, 
Gehrmann-De Ridder, Gehrmann, and Glover have proposed forms for the 
spin-summed antenna splitting functions of massless quarks and 
gluons~\cite{Gehrgluons,Gehr}.   Our previous paper reviews this
latter work and compares the results from our method to theirs.
There is no universal form for antenna splitting functions.  The
behavior of the splitting functions is prescribed in the soft and collinear 
limits but, away from those limits, different expressions are possible,
depending on the framework used in the derivation.  The systematic 
differences between the different proposals are explored in \cite{LP}.

Following the methods of \cite{Gehrgluons,Gehr}, 
splitting functions for massive, spin summed antennae were constructed in 
\cite{massive1,massive2,massive3}.  Again, our expressions agree with
these in having the correct soft and quasi-collinear behavior but 
differ away from these limits.  The addition
of mass greatly complicates both the expressions for the splitting
functions and the precise specification of the  boundaries of phase space.
Because of this, we do not present a detailed comparison to other
massive splitting functions here.

The outline of this paper is as follows:  In Section 2, we will analyze
the case of gluon radiation from an antenna
composed of a massive spin-$\half$ fermion ($Q$) and a massless spin 
$\half$ fermion ($q$) in a configuration of zero helicity. All of the 
new complications that arise when we deal with massive particles can be 
illustrated in this context. We will write expressions for the splitting
functions in terms of spinor products of lightlike vectors associated
with the massive vectors of the particles before and after the splitting.
  In Section 3, we will discuss the 
kinematics of these massive splittings and the evaluation of the 
the spinor product expressions.   

With this introduction, we can go systematically through the various
cases of antennae composed of massive and massless particles.  
In Sections 4 and 5 we will analyze in turn 
the cases of antennae with spin 0 and  spin $\half$
 massive particles recoiling against quarks and gluons
in which the antennae emits another quark or gluon.  In 
Section 6, we discuss the analysis of the  
general case of a pair of massive particles, spin 0 or spin $\half$,
radiating gluons.  In Section 7, we discuss antennae that create a 
pair of massive particles. Section 8 gives some conclusions. 
We collect the complete set of massive antenna splitting functions 
derived in this paper in Appendix A.

\section{The spin zero fermion-quark antenna}

The simplest case of a splitting function with massive particles arises
in the system of a massive and a massless fermion created by a spin 0 
operator.  In this section, we will work out the spin-dependent
splitting functions for this case following the prescriptions in \cite{LP}.
We will then discuss the interpretation of these formulae and their 
comparison to the standard Altarelli-Parisi splitting functions for a 
massive quark~\cite{Catani}.

In \cite{LP}, each case of a spin-dependent splitting is associated with a
gauge-invariant operator that creates the antenna.  For this case, the 
required operator is
\beq
           {\cal O} = \bar Q_L q_R
\eeq{spinzeroop}
where $q$ is an ordinary quark whose mass can be ignored and $Q$ is a 
massive quark.  This operator creates a 2-particle state 
\beq
              Q_L  {\bar q}_L 
\eeq{spinzerostate}
with total spin zero about the production axis.  Antennae with overall
opposite helicity or with antiquarks have the same splitting functions,
by the $P$ and $C$ invariance of QCD.

In \cite{LP}, we wrote the basic formula for final-state
antennae splitting of massless particles in the 
following way:  Notate the splitting as $AB \to acb$, with 
\beq
       (A+B)^2 = s_{AB} = Q^2  \ .
\eeq{sABdefin}
Throughout this paper, for any 4-vectors $i$, $j$, we will define
\beq
           s_{ij} = (i+j)^2 =  m_i^2 + 2 i\cdot j + m_j^2 \ .
\eeq{sijdefin}
Let $z_a$, $z_b$, 
$z_c$ be the momentum fractions of $a$, $b$, and $c$ relative to their
maximum value, 
\beq
       z_a =  {2 Q \cdot q\over Q^2} \ , \  \mbox{etc.} \qquad 
   z_a+z_b+z_c = 2 \ .
\eeq{masslesszdefin}
Then the probability of a splitting is given by 
\beq
  \int d\, \mbox{Prob}  =   N_c{\alpha_s\over 4\pi}({Q\over 2K})
                  \int dz_a dz_b\,
                     {\cal S}(z_a,z_b,z_c)
            \ .
\eeq{basicsplitting}
where $N_c =3 $ is the number of colors in QCD and $K$ is the momentum
of the partons in the center of mass system of the 
original 2-particle antenna. 
 In the massless case, $Q/2K = 1$. 
   The distribution 
${\cal S}$ is the splitting function.  In \cite{LP}, we computed this 
function as the 
ratio of 3- to 2- body amplitudes of an appropriate local operator,
\beq
     {\cal S} = Q^2 
 \biggl|  {\M({\cal O}\to acb)\over \M({\cal O}\to AB)}
                  \biggr|^2 \ .
\eeq{basicS}
This formula is still correct for the massive particle antennae discussed
in this paper.
  We will discuss the kinematics of these
antennae in more detail in Section 3.

In the limit in which $c$ becomes collinear with $a$ or $b$, the antenna
splitting functions reduce to the Altarelli-Parisi functions $P(z)$
that describe
$1\to 2$ splittings. For this limit, the formulae are not as simple in 
the massive case as they are in the
all-massless case.  We will present the explicit formulae and check them
for the spin zero antenna later in this section.

To compute the amplitudes in \leqn{basicS}, 
we use the spinor product formalism for 
massive particles of Schwinn and Weinzierl~\cite{SW}.  For a massless
particle, the states of definite helicity are well-defined and Lorentz
invariant.  For a massive particle, the spin states depend on the 
frame chosen to evaluate them.  In the Schwinn-Weinzierl formalism, a 
massless reference vector $q$ is used to define that frame. The spinors
for an outgoing massive fermion of mass $m$ are written
\beq
    \bar u_L(p) = { [q (p+m)\over  \spb  q{p^\flat} } \qquad 
    \bar u_R(p) = { \langle q (p+m)\over \spa q {p^\flat} }  \ , 
\eeq{massiveus}
where the flatted vector $p^\flat$ is defined by 
\beq
    p^\flat =  p -   { m^2 \over 2 q \cdot p} q 
\eeq{flattedp}
A particularly useful choice for $q$ is the lightlike vector in 
the opposite direction from $p$.  Rotating coordinates so that
\beq
    p = (E, 0, 0, p)  \qquad     \mbox{with} \     E^2 = p^2 + m^2 \ ,
\eeq{pwithz}
let 
\beq
    p^\sharp =  \half {(E+p)} (1,0,0,-1) \ ,
\eeq{psharpdef}
Then if we set $q = p^\sharp$, the flatted vector is
\beq
     p^\flat =  \half {(E+p)} (1,0,0,1) \ .
\eeq{pflatwithz}
This is very convenient.  With this choice of $q$, the 
 spinors defined in \leqn{massiveus} are 
just the usual spinors of definite helicity.   Using the basis of 
Dirac matrices where $\gamma^5$ is diagonal,  it is easy to see 
that \leqn{massiveus} reduces to 
\beqa
    \bar u_L  &=& \pmatrix{ \sqrt{{E-p\over 2}} &
                          \sqrt{{E+p \over 2}} \cr}\otimes  
                                    \pmatrix{ 0& 1 \cr}\CR
    \bar u_R  &=& \pmatrix{ \sqrt{{E+p\over 2}} &
                          \sqrt{{E-p \over 2}} \cr}\otimes  
                                    \pmatrix{  1 & 0\cr }  \ .
\eeqa{definiteus}

Using these conventions, we can easily compute the 2 particle matrix
elements of the operator \leqn{spinzeroop}.  Denote the momenta of the 
initial-state heavy quark and light antiquark as $A$ and $B$, respectively.
Then
\beqa
  \M(Q_L \bar q_L) &=&  { [ q A B\rangle \over [q A^\flat]} =  \spa {A^\flat} B
       \cr
  \M(Q_R \bar q_L) &=&  { m \langle q B\rangle \over \spa{q}{A^\flat}}
\eeqa{firstfirsttwo}
The helicity of the $\bar q$ must be $L$, but the heavy quark
created by \leqn{spinzeroop} could be in either spin state.  However, with the
usual definition of helicity, the production of $Q_R \bar q_L$ from a 
spin 0 operator would be forbidden by angular momentum.  Indeed, when 
we set $q = A^\sharp$,
\beq
    \M(Q_R \bar q_L) \sim    \spa {A^\sharp} B  = 0 \ ,
\eeq{ABzero}
because $A^\sharp$ is a lightlike vector parallel to $B$.  The only nonzero 
matrix element is then
\beq
  \M(Q_L \bar q_L) =  \spa {A^\flat} B \ ;
\eeq{lastMforzero}
this gives the denominator in \leqn{basicsplitting}.  It is convenient
that
\beq
          | \spa{A^\flat}B |^2 =  Q^2 - m^2 =  2QK \ ,
\eeq{squareofzero}
with $K$ as in \leqn{basicsplitting}.

It is straightforward to work out the numerator of \leqn{basicsplitting} 
for the four possible spin states of the 3-particle system $Q g q_L$.
As in \cite{LP}, we label the three final-state momenta as $(a,c,b)$, with 
the emitted particle as $c$.  The results, using a general reference vector
$q$ in \leqn{massiveus}, are
\beqa
\M(Q_L g_L \bar q_L) &=&   - {1\over \spb q c } \biggl\{
      {\spa c \afl \bpa{q Q b} \over s_{ac}- m^2 } + 
           { \bpa{q Q \afl} \over \spb bc } \biggr\}    \CR
\M(Q_L g_R \bar q_L) & =&  
         - {\spa \afl b  \bpa{c Q b} \over \spa bc (s_{ac}-m^2)}
   \CR
\M(Q_R g_L \bar q_L) &=&  - {m \over \spb \afl c \spa q \afl } \biggl\{
      {\spa c q \bpa{\afl Q b} \over s_{ac}- m^2 } + 
           { \bpa{\afl Q q} \over \spb bc } \biggr\} 
   \CR
\M(Q_R g_R \bar q_L) &=&   - {m \spa q b  \bpa{c Q b} \over 
                      \spa bc \spa q \afl  (s_{ac}-m^2)}
\eeqa{Qqspinzero}
We have omitted the overall factor of $(gT^a)$.  When we put $q = \ash$, we
can recognize the simplification
\beq
             \bpa{\ash Q \afl}   = \bpa{\afl Q \ash} = 0   \ .
\eeq{asids}
This follows from the fact that the 4-vector $Q$ 
is a linear combination of the two lightlike vectors $\afl$ and $\ash$.
Now square these expressions and combine with \leqn{squareofzero} to 
evaluate \leqn{basicS}.  This gives
\beqa
    {\cal S}(Q_L g_L\bar q_L) &=&  {Q\over 2K}\biggl|
         { \spa \afl c \bpa{\ash Q b} \over \spb \ash c \bpa{c a c} }\biggr|^2
         \CR 
    {\cal S}(Q_L g_R\bar q_L) &=& {Q\over 2K}\biggl|
         { \spa \afl b \bpa{c Q b}\over \spa b c \bpa{c a c} }\biggr|^2
         \CR
    {\cal S}(Q_R g_L\bar q_L) &=& {m^2 Q\over 2K}\biggl|
         { \spa\ash c \bpa{\afl Q b}\over\spa  \ash \afl \spb \afl c 
                           \bpa{c a c} }\biggr|^2
         \CR
    {\cal S}(Q_R g_R\bar q_L) &=& {m^2Q \over 2K}\biggl|
         { \spa \ash b \bpa{c Q b}\over\spa \ash\afl \spa b c  \bpa{c a c} }
                  \biggr|^2
\eeqa{finalSforzero}

In the all-massless case, we managed to produce
 antenna splitting functions that were
simple rational functions of the $z_a$~\cite{LP}.  Here, the 
antenna splitting functions are more complicated, but not 
excessively so.  The main complications come from the denominators
$(s_{ac} - m^2) = [c a c\rangle$, which do not factorize simply, 
and from the multiple
lightlike vectors needed to characterize the state of the massive quark.
In this case, it is not so difficult to write the splitting functions
in terms of 4-vector products:
\beqa
    {\cal S}(Q_L g_L\bar q_L) &=&  {Q\over K}{s_{\afl c}(2 \ash\cdot Q
             \, b\cdot Q -  \ash \cdot b \, Q^2 )
              \over s_{\ash c}(s_{ac} - m^2)^2 } \CR
    {\cal S}(Q_L g_R\bar q_L) &=&  {Q\over K}{ s_{\afl b}(2b \cdot Q
                \, c\cdot Q -  b \cdot c \, Q^2 )
              \over s_{bc} (s_{ac} - m^2)^2 } \CR
    {\cal S}(Q_R g_L\bar q_L) &=& {m^2 Q\over K}{ s_{\ash c}
                 (2\afl \cdot Q
                \, b\cdot Q -  \afl\cdot b \, Q^2 )
              \over s_{\ash \afl} s_{\afl c}(s_{ac} - m^2)^2 } \CR
    {\cal S}(Q_R g_R\bar q_L) &=& {m^2Q \over K}{ s_{\ash b}
                 (2 b \cdot Q
                \, c \cdot Q -  b\cdot c \, Q^2 )
              \over  s_{\ash \afl} s_{bc} (s_{ac} - m^2)^2 } 
\eeqa{finalSforzerovec}
However, the structure of the expressions is more clearly visible in 
the form \leqn{finalSforzero}.

The expressions \leqn{finalSforzero} contain exact tree-level matrix elements
for the transition of the operator ${\cal O}$ to a three-particle state.
They are correctly used in a parton shower
 for any values of $m/Q$ and $p_T/Q$ among the final-state particles, 
as long as the virtuality at the previous  and successive branchings of the 
shower are well separated from $Q$.  In the all-massless case discussed
in \cite{LP}, we made approximations to the splitting functions valid  
in the soft and collinear limits.  It is less  obvious here which
approximations are appropriate, and, in any case, we did not see 
how to  achieve much 
further simplification. So we will stop at this point for this set of 
splitting fuctions and for all of the massive particle splitting functions
quoted in this paper.

To evaluate expressions of the type of \leqn{finalSforzero},
we find it easiest not to convert the expressions in \leqn{finalSforzero}
into 4-vector products or dimensionless scalars built from these but, rather,
to directly evaluate the spinor brackets.  We will discuss a strategy to 
evaluate these brackets in the next section.

Finally, we must discuss the collinear limits and the connection to the
the Altarelli-Parisi splitting functions.  For the spin zero antennae,
this connection is easiest to discuss for the limit $c \parallel b$, where
only massless particles are involved.  We must still take account of the 
fact that, because $b$ and $c$ recoil against a massive particle, their
maximum momentum is limited.  To account for this, let
\beq
             \tilde z_{b,c} =     {z_{b,c}\over (1 - m^2/Q^2)} \ .
\eeq{tildezdef}
so that $\tilde z_{b}$ and $\tilde z_c$ run from 0 to 1 and, 
in the limit $c\parallel b$, $\tilde z_b + \tilde z_c = 1$.  Then, 
in this collider limit, ${\cal S}$ has the singularity 
\beq
      {\cal S} \sim  \delta_{a,A}{Q^2\over s_{bc}}  P_{B\to c}(\tilde z_c) \ .
\eeq{masslesscoll}
The expressions in \leqn{finalSforzero} satisfy this relation.  The
splitting functions to $Q_R g_{L,R}$ must have no collinear singularity.
This follows from the fact that $[\afl Q b \rangle$ and $\spa \ash b$ vanish
when $b$ becomes opposite to $a$.  The cases of $Q_L g_{L,R}$ do have 
singularities proportional to $s_{\ash c}^{-1}$ and $s_{bc}^{-1}$, with the
correct coefficients to match \leqn{tildezdef}.

In the limit $c\parallel a$, where the $1\to 2$ splitting involves a 
massive particle, the limit is slightly more complicated.
For the  splitting of a massive particle, 
the usual Altarelli-Parisi formula for the 
collinear splitting is conventionally  rewritten as
\beq 
    \int d\, \mbox{Prob} = N_c {\alpha_s\over 2\pi}\int dz \int {dp_T^2\over 
     (p_T^2 + z^2 m^2)}  P(z, p_T)  \ .
\eeq{massiveAP}
We divide the usual expressions for $P(z,p_T)$  by 2 so that these
functions give the contribution from one of the two 
antennae that contribute to a collinear singularity.  Mass-suppressed
terms can contain an additional factor of $ (p_T^2 + z^2m^2)$ in the 
denominator; this is why we have allowed the Altarelli-Parisi
function to depend on $p_T$.  With this formalism, for
$c$ becoming parallel to $a$, 
\beq
        {\cal S}(z_a,z_b,z_c) \to  {Q^2\over s_{ac} - m_A^2} P(\tilde z_c,p_T) 
\eeq{APreduction}
where $s_{ac} = (a+c)^2$. Here again, the parameter $\tilde z_c$ must be 
scaled to equal 1 at its maximum value, as in \leqn{tildezdef}.  For the 
present case in which the $(ac)$ system recoils against a massless
parton, $\tilde z_c = z_c$.

To discuss the limits $c\parallel a$, we first need to recall the 
Altarelli-Parisi functions for splitting of a gluon from a massive
fermion.   The Altarelli-Parisi functions are defined in the 
limit of not only collinear but also high energy emission.  For a 
particle of energy $E$ splitting to particles with transverse momentum
$p_T$ and finite masses $m_i$, these functions
describe the regime $p_T \sim m_i \ll E$. For a splitting
$Q \to g Q$, as we have in this case, the spin-summed splitting
function is~\cite{Catani}
\beq
     P(z) =  { 1 + (1-z)^2\over z} - { m^2 \over a\cdot c} 
\eeq{Catanisplit}   
This expression becomes clearer when it is written as a set of 
spin-dependent Altarelli-Parisi functions.    In the convention
defined by \leqn{massiveAP}, 
\beqa
   P(Q_L \to Q_L g_L) &=&  {p_T^2 \over p_T^2 + z^2 m^2}{1\over z} \CR
   P(Q_L \to Q_L g_R) &=&  {p_T^2 \over p_T^2 + z^2 m^2}{(1-z)^2\over z} \CR
   P(Q_L \to Q_R g_L) &=&  {m^2 \over p_T^2 + z^2 m^2}{z^4\over z} \CR
   P(Q_L \to Q_R g_R) &=&  0
\eeqa{APformassivespin}
The sum of these terms does reproduce \leqn{Catanisplit}.  The placement
of the factors of $z$ implements the {\it dead cone} in which soft radiation
from a massive particle is suppressed within a cone of size $1/\gamma$, 
where $\gamma$ is the boost of the heavy particle~\cite{deadcone,deadtwo}. 

We can now compare the $c\parallel a $ limits of our antenna splitting
functions to \leqn{APformassivespin}.  In the collinear limit,
\beq
s_{ac} - m^2  =    {p_T^2 + z^2 m^2\over z(1-z) } \ .
\eeq{sacrewrite}
Using this formula and the collinear limits of the spinor products, we
find that \leqn{finalSforzero}   does satisfy \leqn{APreduction} with
\leqn{APformassivespin}, up to corrections of relative order $m^2/Q^2$.
  In particular, in the limit $c \parallel a$,
$\ash$ becomes collinear with $b$.  Then the vanishing of $\spa \ash b$ with
no compensatory vanishing in the denominator gives the zero in the last 
line of \leqn{APformassivespin}.

The spin-dependent splitting functions in the remaining sections of this 
paper also satisfy these checks on the collinear limits.  For convenience,
we list the complete set of mass-dependent, spin-dependent Altarelli-Parisi
splitting functions that are needed for these checks in Appendix B.

\section{Kinematics of massive antennae}

The splitting functions computed in the previous section were written 
in terms of spinor products of massless vectors associated with the
massive 4-vectors of the antenna.  One should ask, how are these massless
vectors computed?  A similar question arises in the context of the 
formula \leqn{basicsplitting} for the antenna splitting probability.  This 
equation is easily written down as the ratio of a cross section to produce
a 3-body final state, integrated over 3-body phase space, to the cross 
section to produce a 2-body final state, without a radiated parton, 
integrated over 2-body phase space.  In particular, the integral 
$\int dz_a dz_b$ is an integral over 3-body phase space. One should ask,
what is the boundary of the region of integration for these variables,
and how does one sample points in the interior of this region?

For massless antenna, the answers to these questions are straightforward.
For antenna with both radiators in the final state (FF antennae in the 
notation of \cite{LP}), the complete phase space region is the triangle
\beq
      0 < z_a, z_b < 1  \qquad    z_a + z_b > 1
\eeq{completemassless}
and the region well described by the radiation process $AB\to acb$, with
$c$ soft, is the smaller region  where
\beq
   0 <  z_c < z_a < 1  \quad \mbox{and} \quad 0 < z_c < z_b < 1
\eeq{smallermassless}
To create an additional radiated particle in a state with $N$ massless 
particles, we choose a color-connected pair of particles $AB$, boost
so that $A$ and $B$ are of equal length and back-to-back, choose 
$(z_a, z_b)$ as a random point in the region \leqn{smallermassless},
replace the 2-particle system $AB$ by the chosen 3-particle system
$acb$, and, finally, reverse the boost to bring $acb$ back into the 
original frame.   The corresponding phase space regions and algorithms for
antennae including initial-state particles are described in \cite{LP}.
In this paper, however, we will only discuss final-state showers.

We believe that these 4-vector configurations for massless particles 
provide a good starting point for constructing 4-vector configurations
that include massive particles.  Given a point $\{ \ell_i \}$ 
in the phase space of
of $N$ massless particles, one can obtain a point $\{ k_i \}$
in the phase space of 
$N$ massive particles by rescaling
\beq
   \vec  k_i =  \lambda  \vec \ell_i 
\eeq{vecKtok}
where $\lambda$ obeys
\beq
     \sum_i \hat E_i = E_\CM \ , \quad \mbox{with}  \quad 
             \hat E_i = (|\lambda\vec \ell_i|^2 + m_i^2)^{1/2} \ .
\eeq{lambdaeq}
Conversely, every point of the massive phase space can be constructed
uniquely in this way.   The scale factor $\lambda$ is close to unity
unless one of the massive particles is nonrelativistic.
The relation of the phase space measures for the massive and 
massless variables is~\cite{RAMBO}
\beq
     d \Pi_N(k) =   d\Pi_N(\ell) \cdot  \lambda^{2N-4} \, 
          \prod_i {|\lambda \vec \ell_i|\over E_i}
          \ {\sum_i |\lambda \vec \ell_i|\over \sum_i  
                              |\lambda \vec \ell_i|^2/E_i} \ .
\eeq{psmeasure}
We will refer to the massless vectors $\{\ell_i\}$ as the {\it backbone}
of the massive configuration.

We now have a strategy for the constructing the $N$ particle  phase
space of a parton shower that involves massive particles.  Starting
with a system of 2 massless particles, construct a shower of massless
vectors according to the procedure described above.  In each antenna,
let the
momentum fractions of the (massless)
final particles $a$, $b$ be $w_a$, $w_b$.
Rescale within the antenna by $\lambda$ and use the massless vectors
and this value of $\lambda$ to compute the splitting probabilities.
For example, for the splitting described in the previous section
with particle $a$ massive, the equation for $\lambda$ is 
\beq
         E_a + \lambda (|\vec\ell_b| + |\vec\ell_c|) = Q \ .
\eeq{threelambda}
The splitting probability is given by 
\beq
  \int d\, \mbox{Prob}  =   N_c{\alpha_s\over 4\pi}({Q\over 2K})
    \int dw_a dw_b\cdot \lambda^2 \cdot ({\lambda w_a \over E_a})
                     {\lambda 
            \over |\lambda w_a|^2 Q/2E_a   + \lambda (w_b + w_c) }\ 
                     {\cal S}
            \ .
\eeq{basicsplittingwithw}
To evaluate the splitting function ${\cal S}$ we need the flatted and 
sharped vectors $\afl$ and $\ash$.  The first of these is given by
\beq
        \afl  = \half( E_a/\lambda|\vec \ell_a| +1) \, \lambda\ell_a \ ,
\eeq{makeaf}
and $\ash$ is the massless vector of the same length pointing in the 
opposite direction.   Once the configuration is chosen, the three new
massless vectors are boosted back to the frame of the shower, and we are
ready to generate the next antenna.  When the shower is completed, the 
entire backbone must be rescaled to put the final massive particles on 
shell.  In this prescription, the recoil due to emissions is done locally
in each antenna to the extent that the particles are relativistic, but the
recoil for nonrelativistic massive particles is distributed over the whole
shower. 

There is one more complication that should be discussed.
  For a massless particle, the spin state
is determined by the helicity in a way that is independent of frame.
For a massive particle, a change of frame can rotate the spin.  The 
helicity is preserved by rotations and by boosts along the direction of
motion.  Other boosts, at an angle to the direction of motion, change
the spin orientation.  In the massive particle shower described here, 
we ignore this effect.  In any event, it is unimportant when the 
massive particles are relativistic, and this accounts for 
 most of the radiation from these particles.

\section{Antennae with a massive spin 0 particle}

We are now ready to put together a catalogue of the antenna splitting
functions that describe the emission of quarks and gluons in the 
showering of massive particles.  We begin with the case of a spin 0
massive particle $S$ recoiling against a quark or a gluon.

In the quark case, the antenna is described by an operator
\beq
      {\cal O} =  S^\dagger  \bra{2} q_R
\eeq{opforS}
where $\bra{2}$ is a spin-$\half$ spurion that controls the quark polarization.
Here and in the rest of the paper, we will analyze a subset of the various
discrete choices from which the rest can be derived using the $P$ and $C$ 
symmetries of QCD.  Here, for example, the two cases  
\beq
   S\bar q_L \to S g_L \bar q_L  \ \mbox{and} \ S\bar q_L \to S g_R \bar q_L 
\eeq{twocases}
considered below suffice to provide all of the possble spin-dependent 
splitting functions for $S\bar q \to S g \bar q$ and $S q \to S g q$.

The 2-particle matrix element of the operator \leqn{opforS} is
\beq
       \M({\cal O} \to S \bar q_L)  = \spa 2B \ .
\eeq{zeromatrixtwo}
Then, for the 2-particle antenna $S \bar q_L$ with $S$ moving the $\hat 3$ 
direction, 2 should be a massless fermion moving parallel to $S$. 
In the following,
 we will set $2 = A^\flat$.  This choice follows the methods
used in \cite{LP}.  In that paper, the polarization vectors associated 
with operators ${\cal O}$ with nonzero spin are built from massless 
vectors 1 and 2, chosen in the directions of $B$ and $A$, respectively.
With this choice, the denominator of the expression \leqn{basicS}
 for the splitting function is again evaluated as \leqn{squareofzero}.

The 3-particle matrix elements of \leqn{opforS} are 
\beqa
    \M({\cal O} \to S g_L \bar q_L)  &=& 
 { \langle A^\flat (b+c)ac \rangle 
            \over [cac\rangle \spb bc }  \CR
    \M({\cal O} \to S g_R \bar q_L)  &=& - { \spa{A^\flat}{b} \langle bac]
           \over  [cac\rangle \spa bc }
\eeqa{MforSgq}
Here again, we strip off the factors of $g$ and color matrices. 
The final results are surprisingly compact.

For an antenna containing a massive scalar and gluon, we need to find an
operator that defines an antenna whose initial state includes a gluon
of a definite polarization.  For the antenna with a left-handed gluon,
we may choose~\cite{LP}
\beq
  {\cal O} = {i\over \sqrt{2}} S^\dagger  \bra{2} \bar\sigma \cdot F \ket{2}
\eeq{opforSg}
where
\beq
               \bar \sigma \cdot F = \half \bar\sigma^m \sigma^n F_{mn} \ .
\eeq{sFdef}
This operator projects onto anti-self-dual gauge fields
or left-handed physical gluons.  The corresponding operator 
$\sigma\cdot F$ can be used to define the antenna with an initial 
right-handed gluon.
The two-particle matrix elements of \leqn{opforSg} are
\beq
    \M({\cal O} \to S g_L)  = {\spa 2B }^2 \qquad 
    \M({\cal O} \to S g_R)  = 0  \ .
\eeq{MforSg}
The zero for a $g_R$ is just as one should have expected.  As above, 
we set $2=A^\flat$.

There are two types of  3-particle matrix elements of \leqn{opforSg}.
First, the antenna can radiate a gluon.  The corresponding matrix
elements are 
\beqa
    \M({\cal O} \to S g_Lg_L ) 
         &=& {1\over \spb bc} \biggl[{ {\spa{A^\flat}{b}}^2 [bac \rangle \over 
    [cac\rangle } + 2 \spa{A^\flat}{c} \spa{A^\flat}{b} + { {\spa{A^\flat}{c}}^2 [cab \rangle \over 
    [bab\rangle } \biggr] \CR
    \M({\cal O} \to S g_R g_L)  &=& -  { {\spa{A^\flat}{b}}^2 \langle bac]
           \over  [cac\rangle \spa bc }  \CR
    \M({\cal O} \to S g_L g_R)  &=& - { {\spa{A^\flat}{c}}^2 \langle cab]
           \over  [bab\rangle \spa bc }  \CR
    \M({\cal O} \to S g_R g_R)  &=& 0 \ , 
\eeqa{MforSgg}
following the pattern established in \leqn{MforSgq}.  Second, the 
gluon may split into a quark-antiquark pair.  For this, we need 
the matrix elements
\beqa
    \M({\cal O} \to S \bar q_R q_L ) 
         &=& - { { \spa{A^\flat}{b}}^2\over \spa bc } \CR
    \M({\cal O} \to S \bar q_L q_R ) 
         &=&  { { \spa{A^\flat}{c}}^2\over \spa bc }  \ .
\eeqa{MforSqq}
The splitting functions derived from these matrix elements using 
\leqn{basicS}  are listed
systematically in Appendix A.

\section{Antennae with a massive spin $\half$ particle}

In the same way, we can construct operators that correspond to the 
initial states of antennae involving a massive Dirac fermion $Q$
with a quark or gluon.  The massive fermion can have helicity $\pm\half$.
Because the $Q$ is massive, an initial left-handed $Q$ can flip
over after radiation to a right-handed $Q$, or vice versa.  We have
seen this  already in the special case considered in Section 2.  In this
section, we will recall the results from Section 2 and compare them to 
those of the other three possible antennae of this type.

The antennae with an initial state containing $F$ and a quark can be 
arranged in a state with total spin about the axis of motion $|J^3|$ equal
to 0 or 1.  The spin 0 case was considered in Section 2.  The  
appropriate operator
${\cal O}$ is
\beq  
          {\cal O} = \bar Q q_R \ .
\eeq{OforFq}
The matrix elements of this  operator between two-particle $F\bar q$ 
states are
\beq
    \M({\cal O}\to Q_L \bar q_L) = \spa{A^\flat} B \qquad 
  \M({\cal O}\to Q_L \bar q_R) = 0 
\eeq{MforFqtwo}
in our convention that $A^\sharp$ should be used as  the reference vector
for $Q$.  The three-particle matrix elements are then readily computed.
If we use $\ash$ from the beginnning 
as the reference vector for $Q$, \leqn{Qqspinzero} gives
\beqa
\M(Q_L g_L \bar q_L) &=&   -  
      {\spa c \afl \bpa{\ash Q b} \over \spb \ash c [cac \rangle}\CR
\M(Q_L g_R \bar q_L) & =&  
         - {\spa \afl b  \bpa{c Q b} \over \spa bc [cac\rangle}
   \CR
\M(Q_R g_L \bar q_L) &=&  -m 
    {\spa c \ash \bpa{\afl Q b} \over  \spb \afl c \spa \ash \afl [cac\rangle}
   \CR
\M(Q_R g_R \bar q_L) &=&   -m { \spa \ash b  \bpa{c Q b} \over 
                      \spa bc \spa \ash \afl [cac\rangle } \ .
\eeqa{Qqspinzeroagain}
The antenna splitting function can be constructed from these elements 
in the manner described in Section 2.

The spin 1 case can be treated in the same way.  As described in \cite{LP}
and  at the beginning of Section 4, we introduce lightlike vectors 1 and 2
in the direction of $B$ and $A$, respectively.  Then an appropriate 
operator to define this antenna is
\beq
   {\cal O} =  {\bar Q} \, 1\rangle [ 2\, q_L   \ .
\eeq{OforFqone}
The two-particle matrix elements of this operator are
\beq
    \M({\cal O}\to Q_L \bar q_R) 
 = \spa {A^\flat} 1 \spa 2B \ ,  \qquad 
  \M({\cal O}\to Q_R \bar q_R) = 0 \ ,
\eeq{MforFqtwospinone}
Thus, this operator does correctly represent the initial situation.  We will set
$2=A^\flat$ and $1=B$ in the following expressions.

The splitting function for the antenna to radiate a gluon is computed
from the three-particle matrix elements of this operator to $F g\bar q$
final states.  These are
\beqa
\M(Q_L g_R \bar q_R) &=&   - {\spa {a^\flat} B   [ A^\flat(b+c)ac ] \over \langle c ac ]\spa bc }
       \CR
\M(Q_L g_L \bar q_R) & =&    - {\spb{A^\flat}{b} \over \langle c ac ]\spb bc  \spb {a^\sharp}{a^\flat}}
   \biggl\{  [ a^\sharp a c \rangle [ b Q B\rangle 
            + m^2 { \spa cB \spb {a^\sharp} b}
              \biggr\}    \CR
\M(Q_R g_R \bar q_R) &=&   {m \spa {a^\sharp} B [ A^\flat (b+c)ac ]  \over 
      \spa{a^\sharp} \afl  \langle c ac ]\spa bc }
                   \CR    
\M(Q_R g_L \bar q_R) &=&    {m \spb{A^\flat}{b}   \over 
      \spa{a^\sharp} \afl  \langle c ac ]\spa bc }
      \biggl\{  \spa \ash B \langle cab ]  + \spa {a^\sharp} c 
                \langle B c b ]  \biggr\}   \ .
\eeqa{Qqgspinone}
The splitting functions derived from these formulae and those in 
\leqn{Qqspinzeroagain} are catalogued in Appendix A.
   
For the antennae with $Q$ and a gluon, we again use the operator
$\bar \sigma \cdot F$ to define the initial state as containing a 
gluon of definite left-handed polarization.  There are two cases,  
with total spin $\half$ and $\thalf$.  For the spin $\half$ case,
the appropriate operator is
\beq
     {\cal O} = -{i\over \sqrt{2}} \bar Q \, \bar\sigma \cdot F \ket{2} \ .
\eeq{OforFg}
The dominant two-particle matrix element of this operator is
\beq
    \M({\cal O}\to Q_L \bar g_L) 
 =   \spa {A^\flat} B \spa 2B   \ .
\eeq{MforFgtwo}
If we recall that the vector 2 is identified with $A^\flat$, we see that this
puts the initial $Q$ and $g$ into just the correct orientation.
The matrix elements to $Q_Lg_R$, $Q_Rg_L$, and $Q_Rg_R$ all vanish
if 1 is taken parallel to $B$.

\begin{figure}[t]
\begin{center}
\includegraphics[width=6.0in]{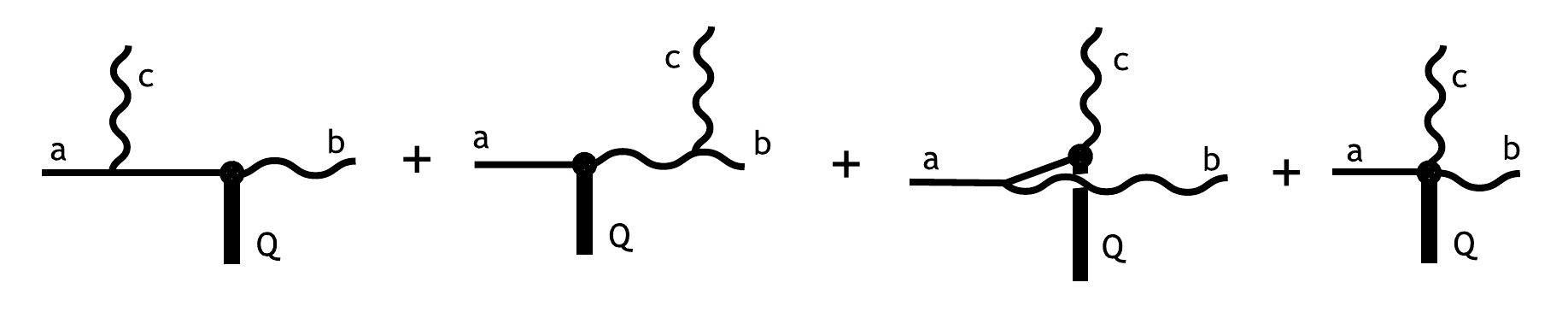}
\caption{Feynman diagrams for the computation of the $Fg\to Fgg$ 
splitting functions~\cite{LP}.}
\label{fig:Fgg}
\end{center}
\end{figure}

The splitting functions for the radiation of a gluon from this antenna
are  given by the matrix elements of \leqn{MforFgtwo} to $Qgg$ final
states.  As in \cite{Gehr} and in \cite{LP}, these matrix elements are 
given by the computation of the set of diagrams shown in Fig.~\ref{fig:Fgg}.
The last diagram in the figure comes from the two-gluon vertex of the
operator $\bar\sigma\cdot F$.  The third diagram is required to make the
computation gauge-invariant.  Its origin is most easily seen by thinking
of the $Q$ as a color octet.  Then this diagram is obviously an essential
contribution to the radiation from the $Q g$ dipole.

With this observation, we find for the three particle matrix elements
of \leqn{MforFgtwo}
\beqa
\M(Q_L g_L  g_L) &=&   
 {1 \over \spb bc \spb \ash \afl}
   \biggl\{ {\spa{A^\flat}{b} \over \langle cac ] }(
   Q^2 \bpa{\ash a c} - m^2 \bpa{\ash Q c}) 
   +
    {\spa{A^\flat}{c} \over \langle bab ] }(
   Q^2 \bpa{\ash a b} - m^2 \bpa{\ash Q b}) 
   \biggr\}\CR
\M(Q_L g_R  g_L) & =&     {\spa {a^\flat} b \spa{A^\flat}{b} \langle b ac ]
         \over  \langle c ac ]\spa bc }    \CR
\M(Q_L g_L  g_R) &=&    {\spa {a^\flat} c \spa{A^\flat}{c} \langle c ab ]
         \over  \langle b ab ]\spa bc }    \CR
\M(Q_L g_R  g_R) & =&  0  \CR
\M(Q_R g_L  g_L) &=&    {m \over \spa{a^\sharp} {a^\flat}\spb bc}
   \biggl\{{ \spa{A^\flat}{b}\over  \langle cac ] } 
    ( \langle {a^\sharp} a Qc \rangle - Q^2  \spa {a^\sharp} c )  +  
    {   \spa{A^\flat}{c} \over  \langle bab ] }
 ( \langle {a^\sharp} a Qb \rangle - Q^2  \spa {a^\sharp} b )
      \biggr\} \CR
 \M(Q_R g_R  g_L) & =&  {m \spa {a^\sharp} b  \spa{A^\flat}{b} \langle b a c]
        \over \spa {a^\sharp}{a^\flat} \langle c ac ]\spa bc } \CR
\M(Q_R g_L  g_R) &=&    {m \spa {a^\sharp} c  \spa{A^\flat}{c} \langle c ab]
        \over \spa {a^\sharp}{a^\flat} \langle b ab ]\spa bc }   \CR
\M(Q_R g_R  g_R) & =& 0 \ .
\eeqa{Qggspinhalf}

The case of a $Qg$ antennae in the spin $\thalf$ state is treated similarly.
The operator that defines the initial state is
\beq
     {\cal O} = -{i\over \sqrt{2}} \bar Q \, 1]\bra 2
                \bar\sigma \cdot F \ket{2} \ .
\eeq{OforFgthalf}
The two-particle matrix elements of this operator are
\beq
    \M({\cal O}\to Q_R \bar g_L) 
 =   \spa 1 {A^\flat}{ \spa 2B}^2  
\eeq{MforFgtwothalf}
and all other matrix elements are equal to zero 
for the choice of 1 parallel to $B$.  We will set $2=A^\flat$ 
and $1=B$ in the expressions that follow.
  
The three-particle matrix elements of \leqn{OforFgthalf} to $Qgg$ final
states are
\beqa
\M(Q_R g_L \bar g_L) &=&   -  { \spb \afl B \over \spb bc}
   \biggl\{  { \spa{A^\flat }{b}  
     \langle A^\flat  (b+c) a c \rangle  \over \langle cac ] }   +
          { \spa{A^\flat}{c}  
     \langle A^\flat (b+c) a b \rangle  \over \langle bab ] } \biggr\}  \CR
\M(Q_R g_R \bar g_L) & =&   - { {\spa{A^\flat}{b}}^2 \over   \langle c ac ]\spa bc }
  \biggl\{  \spb \afl c \langle bQB] + m^2{ \spa \ash b \over \spa \ash \afl } 
         \spb cB \biggr\} \CR
\M(Q_R g_L \bar g_R) &=&   - { {\spa{A^\flat}{c}}^2 \over   \langle b ab ]\spa bc }
  \biggl\{  \spb \afl b \langle cQB] + m^2{ \spa \ash c \over \spa \ash \afl } 
         \spb bB \biggr\} \CR 
         \M(Q_R g_R \bar g_R) &=&   0 \CR 
\M(Q_L g_L \bar g_L) &=&  
       -  {m \spb \ash B \over \spa{a^\sharp} {a^\flat}\spb bc}
   \biggl\{  { \spa{A^\flat}{b}  
     \langle A^\flat (b+c) a c \rangle  \over \langle cac ] }   +
          { \spa{A^\flat}{c}  
     \langle A^\flat (b+c) a b \rangle  \over \langle bab ] }  \biggr\} \CR
 \M(Q_L g_R \bar g_L) & =& - {m { \spa{A^\flat}{b}}^2 
        \over \spb {a^\sharp}{a^\flat} }{(\spb \ash B  \langle bac] + 
     \spb \ash c \langle b c B])\over  \langle c ac ]\spa bc } \CR
\M(Q_L g_L \bar g_R) &=&   - {m { \spa{A^\flat}{c}}^2 
        \over \spb {a^\sharp}{a^\flat} }{(\spb \ash B  \langle cab] + 
     \spb \ash b \langle cb B])\over  \langle b ab ]\spa bc } \CR 
\M(Q_L g_R \bar g_R) & =& 0 \ .
\eeqa{Qggspinthalf}
The splitting functions for $Qg\to Qgg$ that are derived from these
expressions and those in \leqn{Qggspinhalf} are catalogued in Appendix A.

The $Qg$ antennae can also radiate by gluon splitting to a pair of quarks.
For the spin $\half$ case, the relevant matrix elements are
\beqa
    \M(Q_L \bar q_R q_L ) 
         &=&  { \spa \afl b { \spa{A^\flat}{b}}\over \spa bc } \CR
    \M(Q_L \bar q_L q_R ) 
         &=& - {\spa \afl c { \spa{A^\flat}{c}}\over \spa bc }  \CR
    \M(Q_R \bar q_R q_L ) 
         &=&  {m  \spa \ash b { \spa{A^\flat}{b}}\over\spa \ash \afl \spa bc }
 \CR
    \M(Q_R \bar q_L q_R ) 
         &=& - {m \spa \ash c { \spa{A^\flat}{c}}\over\spa \ash \afl \spa bc }
  \CR
\eeqa{MforFqqhalf}
For the spin $\thalf$ case, the matrix elements are
\beqa
    \M(Q_R \bar q_R q_L ) 
         &=&  { \spb\afl B { \spa{A^\flat}{b}}^2\over \spa bc } \CR
    \M(Q_R \bar q_L q_R ) 
         &=& - {\spb \afl B { \spa{A^\flat}{c}}^2\over \spa bc }  \CR
    \M(Q_L \bar q_R q_L ) 
      &=&  {m  \spb \ash B { \spa{A^\flat}{b}}^2\over\spb \ash \afl \spa bc }
                      \CR
    \M(Q_L \bar q_L q_R ) 
   &=& - {m \spb \ash B { \spa{A^\flat}{c}}^2\over\spb \ash \afl \spa bc }  
\eeqa{MforFqqthalf}
The splitting functions for $Qg\to Qgg$ that are derived from these
expressions are catalogued in Appendix A.

\section{Antennae of a pair of massive particles}

After a pair of massive scalars or fermions are produced, their first
emission of a gluon is described by an antenna in which the two massive
particles both appear.  For a complete description, we need the splitting
functions for these  antenna as well.  These formulae are somewhat 
more complicated than those derived above, since some of the simplifications
that are possible when the particle $b$ is  massless  no longer apply.
There is little additional complexity in the cases in which 
the two massive particles have different masses, so we will write the 
formulae for that more general situation.

The case of a pair of scalars is relatively straightforward.  The
scalar particles themselves are spinless, so there is only one case, 
described by the spin-0 operator
\beq
     {\cal O} =   S^\dagger_1 S_2 \ .
\eeq{SSop}
The matrix element of this operator to create the state $S_1\bar {S}_2$ is 
simply 1.  The matrix elements for gluon emission are
\beqa
    \M(S_1 g_L  \bar {S}_2) 
         &=&  {1 \over \spb \afl c }\biggl\{ 
        { [ \afl a c\rangle \over [ cac\rangle } -
          {  \langle \afl b c]\over \langle cbc]} \biggr\} \CR
    \M(S_1 g_R \bar {S}_2 ) 
         &=& - {1\over  \spa \afl c }\biggl\{ 
        { \langle \afl a c]\over \langle cac]} -
          {  \langle \afl b c]\over \langle cbc]} \biggr\} \ .
\eeqa{MforSSg}
Each expression can be brought down to one term using the Schouten 
identity 
\beq
    \apb{ caf} \apb {dbg} - \apb {daf} \apb {cbg} =  - \spa cd [ fabg ] \ .
\eeq{moreSchouten}
This identity is valid when $a$ and $b$ are massive vectors, possibly
with different masses; $c$, $d$,
$f$, and $g$ must be massless.  To prove the identity, write $a$ as a 
linear combination of $\afl$ and $\ash$.  Using \leqn{moreSchouten},
\beqa
    \M(S_1 g_L  \bar {S}_2) 
         &=& - {\langle c ab c \rangle \over \apb{cac} \bpa {cbc} }\CR
    \M(S_1 g_L  \bar {S}_2) 
         &=& {[ c ab c ]\over \bpa{cac} \apb {cbc} } \ .
\eeqa{MforSSgtwo}
The splitting functions are readily assembled from these expressions.

For the antenna of a massive fermion and a massive scalar, the general 
case is described by the spin $\half$ operator
\beq
     {\cal O } =    \bar Q_1\, 1\rangle S_2 \ .
\eeq{OforQSspinzero}
The two-body matrix elements of this operator are
\beq
    \M(Q_{1L} \bar S_2) =    \spa {A^\flat} 1   
\eeq{MforFSspinhalf}
and zero for $Q_{1R}$.  If we take $1 = B^\flat$ following the 
prescriptions above, 
\beq
       |\spa {A^\flat} 1 |^2  =   (E_1+K)(E_2+K) \ ,
\eeq{evalmassiveprods}
where $E_1$, $E_2$, and $K$ are the two energies and the momentum in 
the antenna center of mass frame.

The matrix elements for the operator \leqn{OforQSspinzero}
to create $Q g \bar S$ states is given by the expression
\beq
\M = -{gT^a\over \sqrt{2}} \bar u(a) \biggl[ 
    {\eps \!\! /(c) (a\!\!\! / + c\!\! / + m)\over [cac\rangle} \, 1\rangle - 1\rangle
            { 2 b \cdot \eps(c)\over [cbc\rangle} \biggr] \ ,
\eeq{FSMatrix}
where $\eps(c)$ is the polarization vector of the gluon.  A convenient
way to treat this is to manipulate
\beq
    \eps \!\! /(c)(a\!\!\! / + c\!\! / + m) =  2a\cdot \eps(c) + \eps \!\! / (c) c\!\! / 
\eeq{Asmanip}
plus a term proportional to $(a\!\!\! /-m)$ that gives zero when applied to 
$\bar u(a)$.  The first term in \leqn{Asmanip} combines with the 
last term in \leqn{FSMatrix} to give an amplitude proportional that of
the scalar-scalar case, \leqn{MforSSg} or \leqn{MforSSgtwo} above.
The term with $\eps \!\! /(c)$ vanishes for $g_R$ and gives a simple but 
nonzero term for $g_L$.   The final results for the two amplitudes,
after dropping the factor of $(gT^a)$, are
\beqa
    \M(Q_L g_R  \bar {S}) 
         &=&  {\spa \afl 1 \langle c ab c \rangle 
                                      \over \apb{cac} \bpa{cbc} }\CR
    \M(Q_R g_R  \bar {S}) 
         &=&  {m_1 \spa \ash 1 \langle c ab c \rangle 
               \over \spa \ash \afl \apb{cac} \bpa{cbc} }\CR
    \M(Q_L  g_L  \bar {S}) 
         &=&  -{ \spa \afl 1 [ c ab c ]\over \bpa{cac} \apb{cbc} }
                  + {\spa \afl c \spa c 1 \over \bpa{ cac } } \CR
    \M(Q_R  g_L  \bar {S}) 
         &=& -{m_1\over \spa \ash \afl }
 \biggl[ { \spa \ash 1 [ c ab c ]\over \bpa{cac} \apb{cbc} }
                  - {\spa \ash c \spa c 1 \over \bpa{ cac } }\biggr] \ . 
\eeqa{finalQgS}
Here $m_1$ is the mass of the fermion $Q_1$.  The formulae apply for any 
values of the masses of the fermion and scalar, as long as the 4-vectors
$a$ and $b$ are properly on mass shell.

The decomposition of the gluon coupling to a massive fermion given in
\leqn{Asmanip} is equivalent the representation of this coupling by the
second-order Dirac equation, in which the fermion is replaced by a
field with a scalar-type coupling and a magnetic moment coupling.
The single-gluon magnetic moment coupling has a chiral structure and vanishes for specific combinations of the fermion and gluon spin.  This second-order Dirac formalism is discussed in more detail in~\cite{LPtop}.

For massive fermions, there are two cases, corresponding to total
spin 0 and 1 along the antenna axis.   For the spin 0 case, we 
could use the operator $\bar Q_L Q_L$ to create the antenna, similarly
to the choices in Sections 2 and 5.   However, in the case in which 
both fermions are massive, that operator creates both $Q_L \bar Q_L$ and
$Q_R \bar Q_R$ states.   We will avoid that problem here by taking the
operator that creates an initial state of $Q_L \bar Q_L$ to be
\beq
     {\cal O } =    \bar Q_1\, 1\rangle \langle 2\,  Q_2
\eeq{OforFFspinzero}
The two-body matrix elements of this operator are
\beq
    \M(Q_{1L} \bar Q_{2L}) =    \spa {A^\flat} 1   \spa 2 {B^\flat} \ , 
\eeq{MforFFspinzero}
and zero for the other three helicity states.   Similarly, for the spin
1 case, we will use the operator 
\beq
     {\cal O } =     \bar Q_1\, 1\rangle [ 2 \,  Q_2 \ .
\eeq{OforFFspinone}
to create an initial state of $Q_L \bar Q_R$.
The two-body matrix elements of this operator are
\beq
    \M(Q_{1L} \bar Q_{2R}) =   \spa  {A^\flat} 1 \spb 2 {B^\flat} \ ,
\eeq{MforFFspinone}
and zero for the other three helicity states.  The $Qg\bar Q$ matrix
elements of these operators are easily computed using the methods presented
earlier in this section.  The results for the splitting functions are 
tabulated in Appendix A.

\section{Antennae with massive particle production}

\begin{figure}[t]
\begin{center}
\includegraphics[width=2.0in]{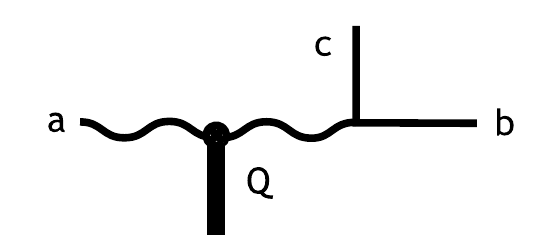}
\caption{The single Feynman diagram for the computation of the $gg\to 
g F \bar F$
splitting function~\cite{LP}.}
\label{fig:pair}
\end{center}
\end{figure}

There is one more situation that we must consider.  At very high energies, 
massive particles can be produced by gluon splitting.   At the LHC, for 
example, parton-parton scattering can give quark-gluon and 
gluon-gluon collisions with center of mass energies well above 1~TeV.
Final state gluon antennae in these collsions can produce pairs of top 
quarks.  The pair production amplitudes are relatively simple, since each
requires only one Feynman diagram, as shown in Fig.~\ref{fig:pair} for the
$gg\to g \bar t  t 
$ case.  The final pair of heavy particles must have 
equal mass and equal spin.  However, there are a large number of cases to 
enumerate.  The massive scalar or fermion pair can be formed from a 
spin $\half$ or a spin $\thalf$ $qg$ antenna or from a spin 0 or 
spin 2 $gg$ antenna.

For scalar pair production, the formalism is actually quite simple.
The spin $\half$ and spin $\thalf$ $qg$ antennae can be represented 
by the operators 
\beqa
     {\cal O}_{1/2} &=& 
        -{i\over \sqrt{2}} \bar q \, \bar\sigma \cdot F \ket{2} \CR 
      {\cal O}_{3/2} &=& 
        -{i\over \sqrt{2}} \bar q \, 1]\bra 2  \bar\sigma \cdot F \ket{2} \ .
\eeqa{opsforSSpair}
If the gluon splits to a pair of scalars, both cases involve the 
operator $\bar \sigma \cdot F$ dotted with the $gS\bar S$ vertex.  This 
product is
\beq
    {1\over 2}[(b\!\!\! /+c\!\! /)\gamma^\mu - 
       \gamma^\mu (b\!\!\! /+c\!\! /) ]   (b-c)_\mu  \equiv
               [b,c]
\eeq{vertexforSS}
so that the relevant three-particle matrix elements are
\beq
   \M(q_L \bar S S) =  - {\langle a [b,c] 2 \rangle \over s_{bc} }
\eeq{SSspinhalf}
for the spin $\half$ case and
\beq
   \M(q_R \bar S S) =   {[a1]\langle 2 [b,c] 2 \rangle \over s_{bc} }
\eeq{SSspinthalf}
 for the spin $\thalf$ case.  

Similarly, the spin 0 and spin 2 
$gg$ antennae, corresponding to the $g_Lg_L$ and $g_Rg_L$ initial states,
 can be represented by the operators
\beqa
     {\cal O}_{0} &=& \half \, \tr [  (\bar \sigma \cdot F)^2 ]  \CR 
      {\cal O}_{2} &=& 
     [ 1|\, \sigma \cdot F \,  |1]\bra2  
              \bar \sigma \cdot F  \ket2  \ .
\eeqa{opsforSSpairg}
The manipulation \leqn{vertexforSS} again gives a simple form for the
three-particle matrix elements.  The corresponding splitting functions
are given in Appendix A.

For the case of massive fermion pair production, this formalism is 
necessarily more complex.  With the choice of helicity states that we 
have used throughout this paper, the vertex to create a pair of massive
fermions is a Dirac matrix.  For the case of a final-state $\bar Q_R Q_L$,
for example, this matrix has the form
\beq
       V = \cfl ]\langle \bfl -  {m^2 \over \spa \csh \cfl \spb \bfl \bsh}
                           \bsh ] \langle \csh
\eeq{VformassiveQ}
Then the matrix element of $\bar\sigma \cdot F$ contains the structure
\beq
          \langle  R_1 \,  [ (b+c), V]  \,  R_2 \rangle
\eeq{commforV}
with a commutator bracketed between reference vectors $R_1$ and $R_2$.
However, the frame-dependent choice of the vectors $\bsh$, $\csh$ makes
it difficult to simplify this expression further.   It is true that
$(b+c) = Q-a$, where $a$ is now massless.  In some cases, we have $R_1 = a$,
in which case the $a$ term cancels. In other cases, we have $R_1 = 2 = \Afl$,
so that the $a$ term vanishes if $a$ is collinear with $A$.  We list 
the full expressions for these splitting functions in Appendix A.

\section{Conclusion}

In this paper, we have provided new materials for the construction of 
parton showers that include massive spin 0 and spin $\half$ particles.
We hope that this formalism we have presented will be useful in describing
the QCD dynamics of the top quark and other heavy particles at LHC.

\appendix

\section{Catalogue of massive antenna splitting functions}

In this appendix, we catalogue the various antenna splitting functions 
for massless particles derived in this paper.  Antenna splitting functions
not listed are equal to cases listed below that are related by the $P$
and $C$ symmetries of QCD.

\subsection{Splitting functions with one massive scalar}

\subsubsection{Spin $\half$ antenna: initial dipole  $S\bar q_L$}

\beqa
{\cal S}(S g_L \bar q_L) &=& {Q \over 2K} 
\biggl| { \langle A^\flat (b+c)ac \rangle 
            \over [cac\rangle \spb bc }  \biggr|^2 \CR
{\cal S}(S g_R \bar q_L) &=& {Q \over 2K} 
\biggl|{ \spa{A^\flat}{b} \langle bac]
           \over  [cac\rangle \spa bc } \biggr|^2
\eeqa{spin0split12}

\subsubsection{Spin 1 antenna: initial dipole $S g_L$}

\beqa
{\cal S}(S g_L  g_L) &=& {1 \over (2K)^2} 
\biggl|{1\over \spb bc} \biggl[{ {\spa{A^\flat}{b}}^2 [bac \rangle \over 
    [cac\rangle } + 2 \spa{A^\flat}{c} \spa{A^\flat}{b} 
+ { {\spa{A^\flat}{c}}^2 [cab \rangle \over 
    [bab\rangle } \biggr] \biggr|^2
\CR
{\cal S}(S g_R  g_L) &=& {1 \over (2K)^2} 
\biggl| { {\spa{A^\flat}{b}}^2 \langle bac]
           \over  [cac\rangle \spa bc }\biggr|^2
\CR
{\cal S}(S g_L  g_R) &=& {1 \over (2K)^2} 
\biggl|{ {\spa{A^\flat}{c}}^2 \langle cab]
           \over  [bab\rangle \spa bc }\biggr|^2
\CR
{\cal S}(S g_R  g_R) &=& 0
\CR
{\cal S}(S \bar q_R  q_L) &=& {1 \over (2K)^2}
 { \bpa{b A^\flat b}^2 \over \bpa{ bcb}}
\CR
{\cal S}(S \bar q_L  q_R) &=& {1 \over (2K)^2} 
 { \bpa{c A^\flat c}^2 \over \bpa{ bcb}}
\eeqa{spin0split1}

Note that the last two expressions are already squared and evaluate to 
values that are real and positive for the case of a  final-state antenna.
For example, $${\cal S}(S \bar q_R  q_L) ={1 \over (2K)^2} {( 2 b\cdot \Afl)^2 \over 
    2 b \cdot c}\ .$$

\subsection{Splitting functions with one massive fermion}

\subsubsection{Spin 0 antenna: initial dipole $Q_L\bar q_L$}

\beqa
    {\cal S}(Q_L g_L\bar q_L) &=&  {Q\over 2K}\biggl|
         { \spa \afl c \bpa{\ash Q b} \over \spb \ash c \bpa{c a c} }\biggr|^2
         \CR 
    {\cal S}(Q_L g_R\bar q_L) &=& {Q\over 2K}\biggl|
         { \spa \afl b \bpa{c Q b}\over \spa b c \bpa{c a c} }\biggr|^2
         \CR
    {\cal S}(Q_R g_L\bar q_L) &=& {m^2Q\over 2K}\biggl|
         { \spa\ash c \bpa{\afl Q b}\over 
                     \spa \ash\afl \spb \afl c \bpa{c a c} }\biggr|^2
         \CR
    {\cal S}(Q_R g_R\bar q_L) &=& {m^2Q\over 2K}\biggl|
         { \spa \ash b \bpa{c Q b}\over 
              \spa \ash\afl \spa b c  \bpa{c a c} }\biggr|^2
\eeqa{spin12split0}

\subsubsection{Spin $\half$ antenna: initial dipole $Q_L g_L$}

\beqa
    {\cal S}(Q_L g_L g_L) &=&  {1\over (2K)^2}\biggl|
    {\spa{A^\flat}{b} (
   Q^2 \bpa{\ash a c} - m^2 \bpa{\ash Q c})  
               \over \spb\ash\afl \langle cac ]  \spb bc }
   +
    {\spa{A^\flat}{c} (
   Q^2 \bpa{\ash a b} - m^2 \bpa{\ash Q b})  
              \over \spb\ash\afl \langle bab ]  \spb bc }
  \biggr|^2
         \CR 
    {\cal S}(Q_L g_R g_L) &=&  {1\over (2K)^2}\biggl|  
     {\spa {a^\flat} b \spa{A^\flat}{b} \langle b ac ]
         \over  \langle c ac ]\spa bc }    
  \biggr|^2
         \CR 
             {\cal S}(Q_L g_L g_R) &=&  {1\over (2K)^2}\biggl|  
             {\spa {a^\flat} c \spa{A^\flat}{c} \langle c ab ]
         \over  \langle b ab ]\spa bc }    
  \biggr|^2
         \CR 
             {\cal S}(Q_L g_R g_R) &=&  0
         \CR 
             {\cal S}(Q_R g_L g_L) &=&  {m^2\over (2K)^2}\biggl|
   { \spa{A^\flat}{b} ( \langle {a^\sharp} a Qc \rangle 
- Q^2  \spa {a^\sharp} c ) \over  \spa\ash\afl \langle cac ]  \spb bc} 
      +  
    {   \spa{A^\flat}{c}  ( \langle {a^\sharp} a Qb \rangle 
- Q^2  \spa {a^\sharp} b ) \over \spa\ash\afl  \langle bab ]\spb bc }
  \biggr|^2
         \CR 
             {\cal S}(Q_R g_R g_L) &=&  {m^2\over (2K)^2}\biggl|    
             { \spa {a^\sharp} b  \spa{A^\flat}{b} \langle b a c]
        \over \spa\ash\afl  \langle c ac ]\spa bc }    
  \biggr|^2
         \CR 
             {\cal S}(Q_R g_L g_R) &=&  {m^2\over (2K)^2}\biggl|    
             { \spa {a^\sharp} c  \spa{A^\flat}{c} \langle c ab]
        \over \spa\ash\afl \langle b ab ]\spa bc }   
  \biggr|^2
         \CR 
             {\cal S}(Q_R g_R g_R) &=& 0
             \CR   
              {\cal S}(Q_L \bar q_R q_L) &=&  {1\over (2K)^2}   
         {\bpa {b \afl b} \bpa{b A^\flat b} \over \bpa {bcb}}
         \CR             
           {\cal S}(Q_L \bar q_L q_R) &=&  {1\over (2K)^2}
      {\bpa {c \afl c} \bpa{c A^\flat c} \over \bpa {bcb}}
         \CR 
           {\cal S}(Q_R \bar q_R q_L) &=&  {m^2\over (2K)^2}
      {\bpa {b \ash b} \bpa{b A^\flat b} \over \bpa{\ash a \ash} \bpa {bcb}}
         \CR 
              {\cal S}(Q_R \bar q_L q_R) &=&  {m^2\over (2K)^2}
      {\bpa {c \ash c} \bpa{c A^\flat c} \over \bpa{\ash a\ash} \bpa {bcb}}
\eeqa{spin12split12}

As in A.1.2, the last four expressions here are already squared and 
evaluate to real, positive values.

\subsubsection{Spin 1 antenna: initial dipole $Q_L\bar q_R$}

\beqa
    {\cal S}(Q_L g_R\bar q_R) &=&  {1\over (2K)^2}\biggl|
        {\spa {a^\flat} B [ A^\flat(b+c)ac ]  \over \langle c ac ]\spa bc }
  \biggr|^2
         \CR 
    {\cal S}(Q_L g_L\bar q_R) &=& {1\over (2K)^2}\biggl|
         {\spb{A^\flat}{b} ([ a^\sharp a c \rangle [ b Q B\rangle 
            + m^2 { \spa cB \spb {a^\sharp} b}) \over \spb\ash\afl \langle c ac ]\spb bc  }
   \biggr|^2
         \CR
    {\cal S}(Q_R g_R\bar q_R) &=& {m^2\over (2K)^2}\biggl|
         { \spa {a^\sharp} B   [ A^\flat (b+c)ac ] \over \spa\ash\afl
       \langle c ac ]\spa bc }
          \biggr|^2
         \CR
    {\cal S}(Q_R g_L\bar q_R) &=& {m^2\over (2K)^2}\biggl|
    {\spb{A^\flat}{b} (\spa \ash B \langle cab ]  + \spa {a^\sharp} c 
                \langle B c b ] )  \over \spa\ash\afl
      \langle c ac ]\spa bc } 
         \biggr|^2
\eeqa{spin12split1}

\subsubsection{Spin $\thalf$ antenna: initial dipole $Q_R g_L$}
\beqa
    {\cal S}(Q_L g_L g_L) &=&  {m^2Q^2\over (2QK)^3}\biggl|
{ \spb \ash B \over \spb\ash\afl \spb bc}
   \biggl\{  { \spa{A^\flat}{b}  
     \langle A^\flat (b+c) a c \rangle  \over \langle cac ] }   +
          { \spa{A^\flat}{c}  
     \langle A^\flat (b+c) a b \rangle  \over \langle bab ] }  \biggr\}
  \biggr|^2
         \CR 
    {\cal S}(Q_L g_R g_L) &=&  {m^2Q^2\over (2QK)^3}\biggl| 
    { { \spa{A^\flat}{b}}^2 (\spb \ash B  \langle bac] + 
     \spb \ash c \langle b c B])\over \spb\ash\afl  \langle c ac ]\spa bc }  
  \biggr|^2
         \CR 
             {\cal S}(Q_L g_L g_R) &=&  {m^2Q^2\over (2QK)^3}\biggl|   
               { { \spa{A^\flat}{c}}^2 (\spb \ash B  \langle cab] + 
     \spb \ash b \langle cb B])\over  \spb\ash\afl \langle b ab ]\spa bc } 
  \biggr|^2
         \CR 
             {\cal S}(Q_L g_R g_R) &=&  0
         \CR 
             {\cal S}(Q_R g_L g_L) &=&  {Q^2\over (2QK)^3}\biggl|
             { \spb \afl B \over \spb bc}
   \biggl\{  { \spa{A^\flat }{b}  
     \langle A^\flat  (b+c) a c \rangle  \over \langle cac ] }   +
          { \spa{A^\flat}{c}  
     \langle A^\flat (b+c) a b \rangle  \over \langle bab ] } \biggr\}
       \biggr|^2
         \CR 
             {\cal S}(Q_R g_R g_L) &=&  {Q^2\over (2QK)^3}\biggl|   
             { {\spa{A^\flat}{b}}^2 \over   \langle c ac ]\spa bc }
  \biggl\{  \spb \afl c \langle bQB] + m^2{ \spa \ash b \over \spa \ash \afl } 
         \spb cB \biggr\} 
  \biggr|^2
         \CR 
             {\cal S}(Q_R g_L g_R) &=&  {Q^2\over (2QK)^3}\biggl| 
             { {\spa{A^\flat}{c}}^2 \over   \langle b ab ]\spa bc }
  \biggl\{  \spb \afl b \langle cQB] + m^2{ \spa \ash c \over \spa \ash \afl } 
         \spb bB \biggr\}    
  \biggr|^2
         \CR 
             {\cal S}(Q_R g_R g_R) &=& 0
                          \CR   
              {\cal S}(Q_L \bar q_R q_L) &=&  {m^2Q^2\over (2QK)^3}   
   {\bpa {B \ash B} \bpa{b A^\flat b}^2 \over \bpa{\ash a\ash} \bpa {bcb}} 
         \CR             
           {\cal S}(Q_L \bar q_L q_R) &=&  {m^2Q^2\over (2QK)^3}
   {\bpa {B \ash B} \bpa{c A^\flat c}^2 \over \bpa{\ash a\ash} \bpa {bcb}} 
         \CR 
           {\cal S}(Q_R \bar q_R q_L) &=&  {Q^2\over (2QK)^3}
     {\bpa {B \afl B} \bpa{b A^\flat b}^2 \over  \bpa {bcb}} 
         \CR 
              {\cal S}(Q_R \bar q_L q_R) &=&  {Q^2\over (2QK)^3}
     {\bpa {B \afl B} \bpa{c A^\flat c}^2 \over  \bpa {bcb}} 
\eeqa{spin12split32}

As in A.1.2, the last four expressions here are already squared and 
evaluate to real, positive values.

\subsection{Splitting functions with two massive scalars}

\subsubsection{Spin 0 antenna: initial dipole $S_1\bar S_2$}

\beqa
    {\cal S}(S_1 g_L  \bar {S}_2) 
   &=&  Q^2 \biggl|{\langle c ab c \rangle \over \apb{cac} \bpa {cbc} }
                   \biggr|^2 \CR
   {\cal S}(S_1 g_R  \bar {S}_2) 
         &=& Q^2  \biggl| {[ c ab c ]\over \bpa{cac} \apb {cbc} }
                    \biggr|^2  
\eeqa{spin0massive}

\subsection{Splitting functions with a massive fermion and a massive scalar}

\subsubsection{Spin $\half$ antenna: initial dipole $Q_1\bar S_2$}

\beqa
    {\cal S}(Q_{1L} g_L  \bar {S}_2) 
   &=&  {Q^2\over (E_1+K)(E_2+K)}
  \biggl|{\spa \afl {B^\flat}\langle c ab c \rangle \over 
                      \apb{cac} \bpa {cbc} }
          - {\spa \afl c \spa c {B^\flat}\over \apb {cac}}
                   \biggr|^2 \CR
    {\cal S}(Q_{1R} g_L  \bar {S}_2) 
   &=& {m_1^2 Q^2\over  (E_1+K)(E_2+K)}
        \biggl|{1\over \spa \ash \afl}\biggl\{ {\spa \ash {B^\flat} 
             \langle c ab c \rangle \over \apb{cac} \bpa {cbc} }
          - {\spa \ash c \spa c \Bfl\over \apb {cac}} \biggr\}
                   \biggr|^2 \CR
    {\cal S}(Q_{1L} g_R  \bar {S}_2) 
         &=& {Q^2 \over  (E_1+K)(E_2+K)} \biggl| {\spa \afl \Bfl
        [ c ab c ]\over \bpa{cac} \apb {cbc} }
                    \biggr|^2   \ .\CR
    {\cal S}(Q_{1L} g_R  \bar {S}_2) 
         &=& {m_1^2 Q^2 \over  (E_1+K)(E_2+K)}
        \biggl| {1 \over \spa \ash \afl}
         {\spa \ash {B^\flat} [ c ab c ]\over \bpa{cac} \apb {cbc} }
                    \biggr|^2   
\eeqa{spinhalf0massive}

\subsection{Splitting functions with two massive fermions}

\subsubsection{Spin 0 antenna: initial dipole $Q_{1L}\bar Q_{2L}$}
\beqa
    {\cal S}(Q_{1L} g_L  \bar {Q}_{2L}) 
   &=&  {Q^2\over ((E_1+K)(E_2+K))^2}
  \biggl|{\spa \afl {B^\flat}\langle c ab c\rangle  \spa {A^\flat}{ b^\flat}
       \over \apb{cac} \bpa {cbc} }\CR & & \hskip 2.0in 
          - {\spa \afl c \spa c \Bfl \spa \Afl \bfl \over \apb {cac}}
          - {\spa \afl \Bfl \spa {A^\flat} c \spa c \bfl \over \bpa {cbc}}
                   \biggr|^2 \CR
    {\cal S}(Q_{1L} g_L  \bar {Q}_{2R}) 
   &=&  {m_2^2 Q^2\over ((E_1+K)(E_2+K))^2}
  \biggl|{1\over \spa {b^\flat}{b^\sharp}}\biggl\{
 {\spa \afl B^\flat\langle c ab c \rangle \spa \Afl \bsh
        \over \apb{cac} \bpa {cbc} }\CR  & & \hskip 2.0in
   - {\spa \afl c \spa c \Bfl \spa \Afl \bsh \over \apb {cac}}
          - {\spa \afl \Bfl \spa \Afl c \spa c \bsh \over \bpa {cbc}}
              \biggr\}     \biggr|^2 \CR
    {\cal S}(Q_{1R} g_L  \bar {Q}_{2L}) 
   &=& {m_1^2 Q^2\over  ((E_1+K)(E_2+K))^2}
        \biggl|{1\over \spa \ash \afl}\biggl\{ {\spa \ash B^\flat 
             \langle c ab c \rangle \spa \Afl \bfl \over \apb{cac} \bpa {cbc} }
      \CR  & & \hskip 2.0in
   - {\spa \ash c \spa c \Bfl \spa \Afl \bfl \over \apb {cac}}
          - {\spa \ash \Bfl \spa {A^\flat} c \spa c \bfl \over \bpa {cbc}}
          \biggr\}   \biggr|^2 \CR
    {\cal S}(Q_{1R} g_L  \bar {Q}_{2R}) 
   &=& {m_1^2 m_2^2 Q^2\over  ((E_1+K)(E_2+K))^2}
        \biggl|{1\over \spa \ash \afl \spa \bsh \bfl}\biggl\{{\spa \ash \Bfl 
             \langle c ab c \rangle \spa \Afl \bsh\over \apb{cac} \bpa {cbc} }
         \CR  & & \hskip 2.0in
   - {\spa \ash c \spa c \Bfl \spa \Afl \bsh \over \apb {cac}}
          - {\spa \ash \Bfl \spa {A^\flat} c \spa c \bsh \over \bpa {cbc}}
                \biggr\}    \biggr|^2 \CR
    {\cal S}(Q_{1L} g_R  \bar {Q}_{2L}) 
   &=&  {Q^2\over ((E_1+K)(E_2+K))^2}
  \biggl|{\spa \afl {B^\flat}[ c ab c]  \spa {A^\flat}{ b^\flat}
       \over \apb{cac} \bpa {cbc} }
                \biggr\}     \biggr|^2 \CR
    {\cal S}(Q_{1L} g_R  \bar {Q}_{2R}) 
   &=&  {m_2^2 Q^2\over ((E_1+K)(E_2+K))^2}
  \biggl|{1\over \spa {b^\flat}{b^\sharp}}
 {\spa \afl {B^\flat} [ c ab c ] \spa \Afl \bsh
        \over \apb{cac} \bpa {cbc} }
                 \biggr|^2 \CR
    {\cal S}(Q_{1R} g_R  \bar {Q}_{2L}) 
   &=& {m_1^2 Q^2\over  ((E_1+K)(E_2+K))^2}
        \biggl|{1\over \spa \ash \afl} {\spa \ash B^\flat 
             [ c ab c] \spa \Afl \bfl \over \apb{cac} \bpa {cbc} }
            \biggr|^2 \CR
    {\cal S}(Q_{1R} g_R  \bar {Q}_{2R}) 
   &=& {m_1^2 m_2^2 Q^2\over  ((E_1+K)(E_2+K))^2}
        \biggl|{1\over \spa \ash \afl \spa \bsh \bfl}{\spa \ash \Bfl 
             [ c ab c ] \spa \Afl \bsh\over \apb{cac} \bpa {cbc} }
                   \biggr|^2
\eeqa{QQspin0massive}

\subsubsection{Spin 1 antenna: initial dipole $Q_{1L}\bar Q_{2R}$}

\beqa
    {\cal S}(Q_{1L} g_L  \bar {Q}_{2L}) 
   &=&  {m_2^2 Q^2\over ((E_1+K)(E_2+K))^2}
  \biggl|{1\over \spa \bfl \bsh}\biggl\{
 {\spa \afl {B^\flat}\langle c ab c\rangle  \spb {A^\flat}{\bsh}
       \over \apb{cac} \bpa {cbc} }
          - {\spa \afl c \spa c \Bfl \spb \Afl \bsh \over \apb {cac}}
             \biggr\}      \biggr|^2 \CR
    {\cal S}(Q_{1L} g_L  \bar {Q}_{2R}) 
   &=&  {Q^2\over ((E_1+K)(E_2+K))^2}
  \biggl|
 {\spa \afl {B^\flat}\langle c ab c \rangle \spb \Afl \bfl
        \over \apb{cac} \bpa {cbc} }
          - {\spa \afl c \spa  c \Bfl \spb \Afl \bfl \over \apb {cac}}
              \biggr\}     \biggr|^2 \CR
    {\cal S}(Q_{1R} g_L  \bar {Q}_{2L}) 
   &=& {m_1^2 m^2_2 Q^2\over  ((E_1+K)(E_2+K))^2}
        \biggl|{1\over \spa \ash \afl \spb \bfl \bsh }
         \biggl\{ {\spa \ash {B^\flat}
             \langle c ab c \rangle \spb \Afl \bsh \over \apb{cac} \bpa {cbc} }
   - {\spa \ash c \spa c \Bfl \spb \Afl \bsh \over \apb {cac}}
          \biggr\}   \biggr|^2 \CR
    {\cal S}(Q_{1R} g_L  \bar {Q}_{2R}) 
   &=& {m_1^2 Q^2\over  ((E_1+K)(E_2+K))^2}
        \biggl|{1\over \spa \ash \afl }\biggl\{{\spa \ash \Bfl 
             \langle c ab c \rangle \spb \Afl \bfl\over \apb{cac} \bpa {cbc} }
   - {\spa \ash c \spa c \Bfl \spa \Afl \bfl \over \apb {cac}}
                \biggr\}    \biggr|^2 \CR
    {\cal S}(Q_{1L} g_R  \bar {Q}_{2L}) 
   &=&  {m_2^2 Q^2\over ((E_1+K)(E_2+K))^2}
  \biggl|{1\over \spb \bfl \bsh}\biggl\{
  {\spa \afl {B^\flat}[ c ab c]  \spb {A^\flat}{\bsh}
       \over \apb{cac} \bpa {cbc} }
          - {\spa \afl \Bfl \spb \Afl c \spb c \bsh \over \bpa {cbc}}
                \biggr\}     \biggr|^2 \CR
    {\cal S}(Q_{1L} g_R  \bar {Q}_{2R}) 
   &=&  {Q^2\over ((E_1+K)(E_2+K))^2}
  \biggl|
 {\spa \afl {B^\flat} [ c ab c ] \spb \Afl \bfl
        \over \apb{cac} \bpa {cbc} }
          - {\spa \afl \Bfl \spb \Afl c \spb c \bfl \over \bpa {cbc}}
                 \biggr|^2 \CR
    {\cal S}(Q_{1R} g_R  \bar {Q}_{2L}) 
   &=& {m_1^2 m_2^2 Q^2\over  ((E_1+K)(E_2+K))^2}
        \biggl|{1\over \spa \ash \afl \spb \bfl \bsh }\biggl\{
 {\spa \ash \Bfl 
             [ c ab c] \spb \Afl \bfl \over \apb{cac} \bpa {cbc} }
          - {\spa \afl \Bfl \spb \Afl c \spb c \bsh \over \bpa {cbc}}
           \biggr\} \biggr|^2 \CR
    {\cal S}(Q_{1R} g_R  \bar {Q}_{2R}) 
   &=& {m_1^2 Q^2\over  ((E_1+K)(E_2+K))^2}
        \biggl|{1\over \spa \ash \afl }\biggl\{ {\spa \ash \Bfl 
             [ c ab c ] \spb \Afl \bfl\over \apb{cac} \bpa {cbc} }
          - {\spa \afl \Bfl \spb \Afl c \spb c \bfl \over \bpa {cbc}}
              \biggr\}     \biggr|^2
\eeqa{QQspin1massive}

\subsection{Splitting functions with pair production of scalars}

\subsubsection{Spin 0 antenna: initial dipole $g_L g_L$}

\beq
    {\cal S}( g_L \bar S S)
   =  {1\over Q^2}
        \biggl|{\langle a [b,c] a \rangle \over s_{bc} }\biggr|^2  \CR
\eeq{SSspinzero}

\subsubsection{Spin $\half$ antenna: initial dipole $q_L g_L$}

\beq
    {\cal S}( q_L \bar S S)
   =  {1\over Q^2}
        \biggl|{\langle a [b,c] A \rangle \over s_{bc} }\biggr|^2  \CR
\eeq{qSSspinhalf} 

\subsubsection{Spin $\thalf$ antenna: initial dipole $q_R g_L$}

\beq
    {\cal S}( q_R \bar S S)
   =  {1\over Q^4}
        \biggl|{\spb a B \langle A [b,c] A \rangle 
                 \over s_{bc} }\biggr|^2  \CR
\eeq{qSSspinthalf} 

\subsubsection{Spin 2 antenna: initial dipole $g_R g_L$}

\beq
    {\cal S}( g_R \bar S S)
   =  {1\over Q^6}
        \biggl|{{\spb a B}^2 \langle A [b,c] A \rangle 
                 \over s_{bc} }\biggr|^2  \CR
\eeq{SSspintwo} 

\subsection{Splitting functions with pair production of fermions}

\subsubsection{Spin 0 antenna: initial dipole $g_L g_L$}

\beqa
   {\cal S}( g_L \bar Q_L Q_L)
   &=&  {m^2\over Q^2 s_{bc}^2 }
        \biggl| {\apb{ a Q \bsh}\spa \cfl a \over \spb \bfl \bsh} 
           +  {\apb{ a Q \csh}\spa \bfl a \over \spb \cfl \csh} \biggr|^2 \CR
{\cal S}( g_L \bar Q_L Q_R)
   &=&  {1\over Q^2 s_{bc}^2 }
        \biggl| \apb{ a Q \cfl}\spa \bfl a 
           + {m^2\over \spb \bfl \bsh \spa \cfl \csh} 
           \apb{ a Q \bsh}\spa \csh a  \biggr|^2 \CR
{\cal S}( g_L \bar Q_R Q_L)
   &=&  {1\over Q^2 s_{bc}^2 }
        \biggl| \apb{ a Q \bfl}\spa \cfl a  
           + {m^2\over \spa \bfl \bsh \spb \cfl \csh} 
           \apb{ a Q \csh}\spa \bsh a  \biggr|^2 \CR
    {\cal S}( g_L \bar Q_R Q_R)
   &=&  {m^2\over Q^2 s_{bc}^2 }
        \biggl| {\apb{ a Q \cfl}\spa \bsh a \over \spb \bfl \bsh} 
          +  {\apb{ a Q \bfl}\spa \csh a \over \spb \cfl \csh} \biggr|^2
\eeqa{QQspinzero}

\subsubsection{Spin $\half$ antenna: initial dipole $q_L g_L$}

\beqa
    {\cal S}( q_L \bar Q_L Q_L)
   &=&  {m^2\over 4Q^2 s_{bc}^2 }
        \biggl| {\apb{ a Q \bsh}\spa \cfl A \over \spb \bfl \bsh} 
           +  {\apb{ a Q \csh}\spa \bfl A \over \spb \cfl \csh} \CR 
   & & \hskip 1.0in +
 {\apb{ A (Q-a) \bsh}\spa \cfl a \over \spb \bfl \bsh} 
           +  {\apb{ A (Q-a) \csh}\spa \bfl a \over \spb \cfl \csh} \biggr|^2
                    \CR
{\cal S}( q_L \bar Q_L Q_R)
   &=&  {1\over 4Q^2 s_{bc}^2 }
        \biggl| \apb{ a Q \cfl}\spa \bfl A 
           + {m^2\over \spb \bfl \bsh \spa \cfl \csh} 
           \apb{ a Q \bsh}\spa \csh A \CR
    & & \hskip 1.0in  + \apb{ A (Q-a) \cfl}\spa \bfl a 
           + {m^2\over \spb \bfl \bsh \spa \cfl \csh} 
           \apb{ A (Q-a) \bsh}\spa \csh a \biggr|^2 \CR
{\cal S}( q_L \bar Q_R Q_L)
   &=&  {1\over 4 Q^2 s_{bc}^2 }
        \biggl| \apb{ a Q \bfl}\spa \cfl A  
           + {m^2\over \spa \bfl \bsh \spb \cfl \csh} 
           \apb{ a Q \csh}\spa \bsh A  \CR
    & & \hskip 1.0in  +  \apb{ A (Q-a) \bfl}\spa \cfl a 
           + {m^2\over \spa \bfl \bsh \spb \cfl \csh} 
           \apb{ A  (Q-a) \csh}\spa \bsh a
                \biggr|^2 \CR
    {\cal S}( q_L \bar Q_R Q_R)
   &=&  {m^2\over 4 Q^2 s_{bc}^2 }
        \biggl| {\apb{ a Q \cfl}\spa \bsh A \over \spb \bfl \bsh} 
           +  {\apb{ a Q \bfl}\spa \csh A \over \spb \cfl \csh} 
          \CR 
   & & \hskip 1.0in + {\apb{ A (Q-a) \cfl}\spa \bsh a \over \spb \bfl \bsh} 
           +  {\apb{ A (Q-a) \bfl}\spa \csh a \over \spb \cfl \csh}
                       \biggr|^2
\eeqa{QQspinhalf}

\subsubsection{Spin $\thalf$ antenna: initial dipole $q_R g_L$}

\beqa
    {\cal S}( q_R \bar Q_L Q_L)
   &=&  {m^2\over Q^4 s_{bc}^2 }
        \biggl|
    \spb a B \biggl\{   
       { \apb{ A(Q-a) \bsh}\spa \cfl A \over \spb \bfl \bsh} 
           +  {\apb{ A (Q-a) \csh}\spa \bfl A \over \spb \cfl \csh}\biggr\}
               \biggr|^2 \CR
{\cal S}( q_R \bar Q_L Q_R)
   &=&  {1\over Q^4 s_{bc}^2 }
        \biggl|
    \spb a B \biggl\{  \apb{ A (Q-a) \cfl}\spa \bfl A 
           + {m^2\over \spb \bfl \bsh \spa \cfl \csh} 
           \apb{ A (Q-a) \bsh}\spa \csh A \biggr\} \biggr|^2 \CR
{\cal S}( q_R \bar Q_R Q_L)
   &=&  {1\over Q^4 s_{bc}^2 }
        \biggl|
    \spb a B \biggl\{    \apb{ A (Q-a) \bfl}\spa \cfl A  
           + {m^2\over \spa \bfl \bsh \spb \cfl \csh} 
           \apb{ A (Q-a) \csh}\spa \bsh A \biggr\} \biggr|^2 \CR
    {\cal S}( q_R \bar Q_R Q_R)
   &=&  {m^2\over Q^4 s_{bc}^2 }
        \biggl|
    \spb a B \biggl\{   
    {\apb{ A (Q-a) \cfl}\spa \bsh A \over \spb \bfl \bsh} 
           +  {\apb{ A (Q-a) \bfl}\spa \csh A \over \spb \cfl \csh} 
                       \biggr\} \biggr|^2
\eeqa{QQspinthalf}

\subsubsection{Spin 2 antenna: initial dipole $g_R g_L$}

\beqa
    {\cal S}( g_R \bar Q_L Q_L)
   &=&  {m^2\over Q^6 s_{bc}^2 }
        \biggl|
    {\spb a B}^2 \biggl\{   
        {\apb{ A (Q-a) \bsh}\spa \cfl A \over \spb \bfl \bsh} 
           +  {\apb{ A (Q-a) \csh}\spa \bfl A \over \spb \cfl \csh}\biggr\}
               \biggr|^2 \CR
{\cal S}( g_R \bar Q_L Q_R)
   &=&  {1\over Q^6 s_{bc}^2 }
        \biggl|
    {\spb a B}^2 \biggl\{    \apb{ A (Q-a) \cfl}\spa \bfl A 
           + {m^2\over \spb \bfl \bsh \spa \cfl \csh} 
           \apb{ A (Q-a) \bsh}\spa \csh A \biggr\} \biggr|^2 \CR
{\cal S}( g_R \bar Q_R Q_L)
   &=&  {1\over Q^6 s_{bc}^2 }
        \biggl|
    {\spb a B}^2 \biggl\{    \apb{ A (Q-a) \bfl}\spa \cfl A  
           + {m^2\over \spa \bfl \bsh \spb \cfl \csh} 
           \apb{ A (Q-a) \csh}\spa \bsh A \biggr\} \biggr|^2 \CR
    {\cal S}( g_R \bar Q_R Q_R)
   &=&  {m^2\over Q^6 s_{bc}^2 }
        \biggl|
    {\spb a B}^2 \biggl\{   
    {\apb{ A (Q-a) \cfl}\spa \bsh A \over \spb \bfl \bsh} 
           +  {\apb{ A (Q-a) \bfl}\spa \csh A \over \spb \cfl \csh} 
                       \biggr\} \biggr|^2
\eeqa{QQspintwo}

\section{Spin-dependent Altarelli-Parisi functions for massive particles}

In this Appendix, we present the spin-dependent Altarelli-Parisi splitting
functions for massless and massive particles.   The massless cases were
derived in the original paper of Altarelli and Parisi~\cite{AP}.  
Spin-summed Altarelli-Parisi functions for the cases with massive
particles arise in NLO QCD calculations for supersymmetric particle
production. They have been catalogued by Catani, Dittmaier, and 
Tr\'ocs\'anyi in \cite{Catani}.  The  spin-dependent functions
can be worked out by textbook methods.  Here we present these functions
 in a representation convenient for comparison to the antenna
splitting functions derived in this paper.  We omit the overall color
factor of $N_c$ and divide by 2 so that the splitting accounts the contents
of an individual antenna.

Note that, since we work at the leading order in $N_c$ and normalize to a
single antenna, there is no difference between the splitting function for a
heavy quark or a gluino to radiate a gluon.  Thus, there are only two
cases, the cases of a heavy scalar $S$ or a heavy quark $Q$
radiating a gluon.  The cases of a heavy particle splitting to 
a heavy particle by radiating  a gluon are given by the same expressions with
$z \to (1-z)$.

For $S \to gS$,
\beqa
   P(S \to S g_L S) &=&  {p_T^2 \over p_T^2 + z^2 m^2}{1-z\over z} \CR
   P(S \to S g_R S) &=&  {p_T^2 \over p_T^2 + z^2 m^2}{1-z\over z} \CR
\eeqa{APformassivezero}

For $Q \to gQ$,
\beqa
   P(Q_L \to Q_L g_L) &=&  {p_T^2 \over p_T^2 + z^2 m^2}{1\over z} \CR
   P(Q_L \to Q_L g_R) &=&  {p_T^2 \over p_T^2 + z^2 m^2}{(1-z)^2\over z} \CR
   P(Q_L \to Q_R g_L) &=&  {m^2 \over p_T^2 + z^2 m^2}{z^4\over z} \CR
   P(Q_L \to Q_R g_R) &=&  0
\eeqa{APformassivehalf}

\Acknowledgements

\noindent The authors thank Kassa Betre, Stefan Hoeche, and Jared Kaplan for 
helpful conversations.
  This work is supported by the US Department of 
Energy under contract DE--AC02--76SF00515. A.L. is also supported
by an LHC Theory Initiative Travel Award.

\end{document}